\documentclass[twocolumn,prb,aps,amsfonts]{revtex4-1}
\usepackage{latexsym}
\usepackage{dcolumn}
\usepackage[dvips]{graphicx}
\usepackage{amssymb}
\usepackage{graphicx}
\usepackage{psfrag, color}
\usepackage{verbatim}
\usepackage{amsmath}
\usepackage{epsf}

\newcommand{\vk}{{\bf k}}

\begin{document}

\draft

\title{Velocity renormalization and anomalous quasiparticle dispersion
  in extrinsic graphene} 
\author{S. Das Sarma$^1$ and E. H. Hwang$^{1,2}$}
\address{Condensed Matter Theory Center, 
$^1$Department of Physics, University of Maryland, College Park,
Maryland  20742-4111 \\
$^{2}$SKKU Advanced Institute of Nanotechnology, Sungkyunkwan
University, Suwon, 440-746, Korea} 
\date{\today}

\begin{abstract}
Using many-body diagrammatic perturbation theory we consider carrier
density- and substrate-dependent many-body renormalization of doped
or gated graphene induced by Coulombic electron-electron interaction
effects. We quantitatively calculate the many-body spectral function,
the renormalized quasiparticle energy dispersion, and the renormalized
graphene velocity using the leading-order self-energy in the
dynamically screened Coulomb interaction within the ring diagram
approximation.  We predict experimentally
detectable many-body signatures, which are enhanced as the carrier
density and the substrate dielectric constant are reduced, finding an
intriguing instability in the graphene excitation spectrum at low
wave vectors where interaction completely destroys all particle-like
features of the noninteracting linear dispersion.
We also make experimentally relevant quantitative predictions about
the carrier density and wave-vector dependence of graphene velocity
renormalization induced by electron-electron interaction. 
We compare on-shell and off-shell self-energy approximations within
the ring diagram approximation, finding a substantial quantitative
difference between their predicted velocity renormalization
corrections in spite of the generally weak-coupling nature of
interaction in graphene.
\end{abstract}

\maketitle

\section{introduction}

Graphene is a unique electronic material which consists simply of a
single two-dimensional (2D) atomic membrane of carbon atoms in a
honeycomb lattice. The seminal discovery \cite{novoselov2004} that
graphene can be gated 
(i.e., ``doped'' with a finite free carrier density) to produce a
variable carrier density ($n$) electron 
or hole 2D system simply by tuning an external gate voltage has given
rise to a truly amazing level of intellectual activity with more that
10,000 graphene publications appearing in the scientific literature
since 2005 covering many disparate fields including physics, chemistry,
materials science, electrical engineering, and chemical
engineering. From the fundamental perspective of quantum condensed
matter physics graphene is an intriguing system \cite{dassarma2011} in
many ways: it is a 
true 2D stable crystal with unique elastic and mechanical properties,
and gated graphene is an exotic 2D gapless and chiral electron system
with linear energy dispersion, i.e., the electron-hole bands in
graphene have Dirac-Weyl massless linear dispersion instead of the
usual parabolic energy dispersion one sees in most solid state
materials. Graphene is thus both a semimetal and a gapless
semiconductor, which can be easily doped by an external gate voltage
to induce a variable carrier density $n$ (either electrons or holes) with
a special charge neutrality point ($n\equiv 0$), often called the
Dirac point, separating electron bands from the hole bands.

The purpose of the current work is to theoretically consider
electron-electron interaction effects in gated graphene (i.e. $n \neq
0$), sometimes also called ``extrinsic'' graphene
\cite{hwang2007a,dassarma2007} to contrast with
very special situation of ``intrinsic'' graphene ($n\equiv 0$) where
the chemical potential is precisely located at the Dirac point. Since
the Dirac point itself ($n=0$ precisely) is by definition a set of
measure zero (particularly because of the gaplessness of graphene
bands), most experiments can only probe extrinsic graphene (with
$n\neq 0$) with any conclusion about the Dirac point being obtained
indirectly through an extrapolation to zero density. The physics of
graphene Dirac point is interesting in its own right as it is
well-established \cite{dassarma2007,gonzalez1994,khvesh2001} that intrinsic
graphene ($n=0$) is not a Fermi 
liquid. The presence of disorder-induced electron-hole
puddles\cite{dassarma2011} makes it difficult to study the physics of
pristine Dirac point physics. 
Interaction physics at the Dirac point is outside the scope of the
current work.

Since our interest is extrinsic gated graphene at $T=0$, we consider
our starting noninteracting system to have the Fermi energy ($E_F$, or
equivalently the chemical potential $\mu$) in the conduction band
(with no loss of generality) given by $E_F = v_0 (4\pi n/g)^{1/2}$,
choosing $\hbar =1$ throughout, where $g=4$ is the graphene ground
state degeneracy arising from spin (2) and valley (2)
degeneracies. The Fermi wave vector $k_F= (\pi n)^{1/2}$ defines the
2D momentum upto which the conduction band with linear noninteracting
band dispersion $E_0(k) = v_0 k$, with $k=|\vk|$ being the 2D wave
vector, is filled. The linear band dispersion leads to a simple linear
$E$ density of states, which then provides $k_F$, $E_F$,
{\it etc.} quoted above \cite{dassarma2011}. Here $v_0 \approx 10^8$
cm/s is the noninteracting 
graphene velocity obtained, for example, from band structure
calculations \cite{dassarma2011}. Electron-electron interaction would
affect the graphene 
noninteracting dispersion $E_0(k) = v_0 k$, perhaps modifying the
dispersion itself, i.e., changing the constant $v_0$, which is
independent of both the carrier density and the substrate material
supporting the graphene layer in the noninteracting single-particle
approximation, to a renormalized wave vector-dependent nonlinear
graphene velocity $v^*(k)$ which depends not only on $k$, but also on
the carrier density $n$ and the substrate material (through the
background dielectric constant $\kappa$).

The main goal of our work is
to make precise (and experimentally verifiable) quantitative
predictions about the interacting quasiparticle energy dispersion and
the renormalized interacting velocity $v^*$ in gated extrinsic
graphene which, in principle, may depend on wave vector, carrier
density, and the substrate material. Quite surprisingly, this issue,
in spite of its considerable importance, has not been studied in
details in the existing theoretical literature as earlier work on the
subject has concentrated mostly on interaction effects in intrinsic
graphene \cite{dassarma2007,gonzalez1994,khvesh2001,elias2011} and
the associated instabilities of the Dirac point or on the 
numerical calculations of the interacting spectral function in a
narrow range of momentum and frequency \cite{polini2008,hwang2008}.
In addition, we investigate the renormalized quasiparticle energy
dispersion for different interaction strength (i.e. $r_s$) values to
investigate instability at low energy regime (i.e., near the Dirac point).
It is expected that the quasiparticle feature may be modified
due to the many body effects (plasmonic feature) as seen in the
ARPES\cite{arpes,arpes1,arpes2}. 
However, unexpectedly we find that the quasiparticle features disappear
as the interaction strength increases.
In earlier work\cite{arpes} some of the features presented here
are discussed for small $r_s$ values ($r_s < 0.5$). However, the
instability due to many body effect is significantly enhanced for
large $r_s$ and for 
low densities, which is not discussed elsewhere.       
This is a very interesting prediction and can be observed
in samples with high $r_s$ value (for example, suspended graphene).
In fact, we find that a careful experimental study of even the standard
graphene on substrates should be able to discern our predicted
signature of a dispersion instability in the spectral function.

We emphasize that there have been several theoretical investigations of
interaction effects in
graphene\cite{dassarma2011,hwang2007a,dassarma2007,gonzalez1994,khvesh2001,elias2011,polini2008,hwang2008,arpes,arpes2}
including some by us \cite{hwang2007a,dassarma2007,hwang2008} as well
as a few careful experimental studies \cite{elias2011,arpes,arpes1,arpes2}
comparing with the theoretical predictions. Our work is based on
existing
\cite{dassarma2011,hwang2007a,dassarma2007,polini2008,hwang2008} and
well-established theoretical techniques (namely, Feynman-Dyson
diagrammatic perturbation theory using the dynamically screened Coulomb
interaction as the effective coupling constant -- a technique that
goes back to the 1960s and has been used extensively and successfully
to calculate many-body interaction effects on single-particle
properties in metals, doped semiconductors, and 2D electron systems as
well as in band structure calculations under the name of '$GW$
approximation'). In spite of the existing theoretical literature
dealing with the calculation of electron-electron interaction-induced
many-body renormalization effects in doped graphene, the results
presented in the current work are completely new and are directly
relevant to experimental measurements. Although based on the standard
$GW$-type theoretical approximation our numerical results for the
coupling constant dependent graphene Fermi velocity as well as the
wave vector dependent graphene velocity are unavailable in the
literature. These quantities can be measured directly by STM/STS or
ARPES measurements to be compared with our theoretical
results. Similarly, our results for the anomalous quasiparticle
dispersion and the associated collapse of the quasiparticle picture
for graphene at low carrier densities are new although the graphene
spectral function itself has earlier been calculated in the literature
\cite{polini2008,hwang2008,arpes,arpes1,arpes2}. Finally, our comparison
between on-shell and off-shell self-energy approximations for graphene
many-body renormalization has not earlier been investigated in the
graphene literature where all earlier work specifically considered the
off-shell solution of the Dyson equation.


The rest of the paper is organized as follows. In Sec. II we discuss
the general theory of the self-energy of graphene. In Sec. III we
present our calculated renormalized Fermi velocities within on-shell and
off-shell approximations. In Sec. IV we
show the detail results of the quasiparticle spectral function and the
instability of the graphene dispersion, and finally we conclude in
Sec. V.


\begin{figure}[t]
	\centering
	\includegraphics[width=1.\columnwidth]{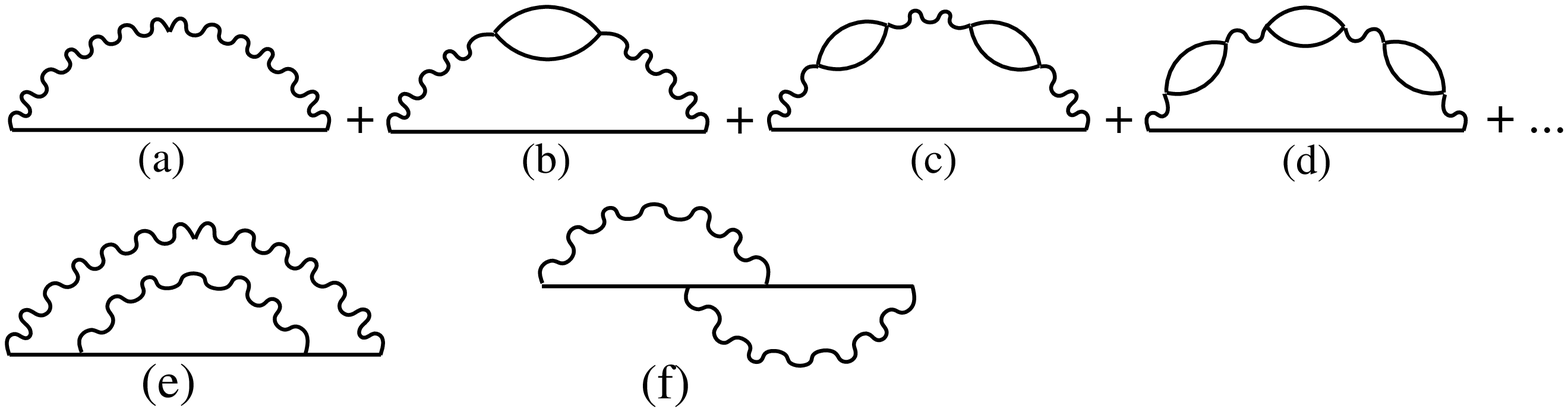}
	\caption{Self-energy contribution [(a)-(d) plus all
            higher-order ring diagrams] included in our theory with the
          wiggly and the non-wiggly lines representing the bare
          interaction ($V$) and Green's function ($G_0$) respectively. Figures (e)
          and (f) represent two second-order contributions ignored in
          the theory, which may approximately cancel each other
          leading to the on-shell approximation being quantitatively
          ``better'' than the off-shell Dyson's solution.
}
\label{fig:fig1}
\end{figure}

\section{self-energy}

The central theoretical quantity we calculate is the graphene
dynamical self-energy function $\Sigma(k,E)$, which, in general,
depends both on momentum (wave vector) and energy (frequency)
independently. An exact theoretical evaluation of the self-energy for
electrons interacting via the long-range Coulomb interaction ($V$) is,
of course, impossible for arbitrary carrier density and interaction
strength, but an excellent approximation scheme, which goes by the
various nomenclatures such as the $GW$-approximation or the dynamical
RPA, is well-established for parabolic metallic 2D and 3D electron
systems. This approximation involves calculating the self-energy in an
infinite order diagrammatic perturbation theory including all diagrams
containing (to all orders) the bare Coulomb interaction, the bare
electron Green's function, and the bare polarizability (i.e., the
bubble or the ring diagrams formed by the closed loop of an electron
and a hole propagator). Equivalently, this infinite series of ring
diagrams (we show in Fig.~\ref{fig:fig1} the self-energy diagrams of the theory)
corresponds to the leading-order perturbation expansion in the
dynamically screened Coulomb interaction ($W$) where the screening is
done by the electronic dynamical dielectric function ($\epsilon$)
obtained in the random phase approximation (RPA) involving the infinite
series of ring diagrams \cite{hwang2007}. 

The validity of considering RPA type diagrams is based on the infrared
divergence of the ring diagrams. All non-RPA diagrams are negligible
compared with the same order RPA diagram. The divergence can be
controlled by the RPA re-summation as shown in Fig.~\ref{fig:fig1}. 
The justification for why non-RPA diagrams in graphene can
be neglected is discussed in Ref.~\onlinecite{dassarma_un}.
We also emphasize that
there is no known controlled technique for systematically calculating
quasiparticle properties in interacting electron systems 
other than the $GW$ approximation based on RPA.  
Within the leading-order dynamical RPA the self-energy $\Sigma(k,E)$
of graphene is given by\cite{mahan,hedin} 
\begin{eqnarray}
\Sigma_s(k,i\omega_n) =
-\frac{1}{\beta}\sum_{s'}\sum_{{q},i\nu_n} 
G_{0,s'}(k+q,i\omega_n+i\nu_n) \nonumber \\
\times W(q,i\nu_n)
F_{ss'}({k},{k}+{q}),
\label{sigma}
\end{eqnarray}
where $\beta = 1/k_BT$ is the inverse temperature, $W$ is the screened
Coulomb interaction,
$G_{0} (k,i\omega_n)=[i\omega_n-E_0(k)+\mu]^{-1}$
is the bare Green's function (where $\omega_n,\nu_n$, and $\mu$ are
Matsubara fermion, boson  
frequencies and chemical potential, respectively)\cite{mahan}. 
Due to the chiral property of graphene the self-energy in
Eq.~(\ref{sigma}) has an additional term (i.e., $F_{ss'}$) which does
not exist in the 
non-chiral systems. The $F_{ss'}$ arises from the overlap of
$\mathinner|\!s{k}\rangle$ and $\mathinner|\!s'{k}'\rangle$ and is
given by $F_{ss'}({k},{k}')
= \frac{1}{2}(1 + ss' \cos\theta_{{kk}'})$, where  
$s,s'=\pm 1$ are band indices of
graphene and $\theta_{{kk'}}$ is the angle between ${k}$, ${k}'$.
The screened Coulomb interaction $W(q,i\nu_n) =
V(q)/\epsilon(q,i\nu_n)$,  where $V(q)=2\pi e^2/\kappa q$ is
the bare Coulomb potential  
($\kappa = $background dielectric constant), and
$\epsilon(q,i\nu_n)$ is the 2D dynamical dielectric function. 
In the RPA $\varepsilon(q,i\nu_n) = 1-V(q)
\Pi_0(q,i\nu_n)$, 
where $\Pi_0=G_oG_0$  is the noninteracting polarizability\cite{hwang2007}.
The expression for $\Sigma$, $W$, $\Pi_0$ involve
multidimensional complex integrals over internal momentum and energy.


Since there has been substantial earlier work in the literature on the
theory of the electron self-energy within the dynamical RPA $GW$
approximation in systems as different as 3D electron gas \cite{hedin}, 
2D electron gas \cite{jalabert}, 1D electron gas \cite{hu1993},
monolayer graphene \cite{polini2008,hwang2008,leblanc,dassarma_un}, and bilayer
graphene \cite{sensarma2011}, we refrain from giving more
theoretical details on our calculations, concentrating instead on the
results and their experimental implications for observable many-body
effects in gated graphene.

We note that ($i$) the calculation of the self-energy $\Sigma$ leads
immediately to the interacting Green's function $G$, [i.e. $G^{-1} =
  G_0^{-1}-\Sigma$] which contains all
information about the single-particle properties of graphene, and that
(ii) once the bare polarizability $\Pi_0$, which was earlier obtained
\cite{hwang2007}, is known, the calculation of self-energy
     [Eq.~\ref{sigma}] would lead 
to $G$ since $G_0$ and $V$ are known by definition. 
We note that we are utilizing the leading-order ``$G_0W$" rather than the
self-consistent ``$GW$" approximation in our theory. We believe the
$G_0W$ approximation to be more meaningful from a diagrammatic
perturbative sense since it is a true leading-order expansion in the
effective RPA screened dynamical interaction $W$ whereas the iterative
$GW$ approximation mixes in unwanted higher-order terms in $W$ which are
inconsistent with the leading-order expansion in $W$.

We 
emphasize that the graphene self-energy
calculation is highly nontrivial and extremely subtle because one must
include both intraband and interband contributions in the
theory. Remembering that the noninteracting Green's function $G_0 =
[E-E_0(k)]^{-1}$ has its poles at the noninteracting energy $E_0 = v_0
k$, we get the following equation for the poles of the interacting
Green's function $E=E_0 + \Sigma(k,E)$, which defines a general
integral equation (Dyson equation) for the interacting graphene
dispersion:
\begin{equation}
E(k) = E_0(k) + {\rm Re}\Sigma[k,E(k)].
\label{eq:dyson}
\end{equation}
We note that $\Sigma(k,E)$ is, in general, complex and the
quasiparticle energy (damping) for the interacting system is given by
the real (imaginary) part of the self-energy. We solve the integral
equation defined by Eq.~(\ref{eq:dyson}) numerically iteratively to obtain the
interacting quasiparticle dispersion $E(k) = v_1^*(k)k$, and refer to
this iterative solution as the ``off-shell'' approximation which
alludes to the full solution of the Dyson equation. An alternative
approximation for the self-energy, the so-called on-shell
approximation, goes back to the early days of diagrammatic many-body
theory \cite{dubois1959}, and has, in particular, been emphasized by
Rice \cite{rice1965}. In the
on-shell approximation, one simply takes the first iterative solution
of Eq.~(\ref{eq:dyson}) to get
\begin{equation}
E(k) = E_0(k) + {\rm Re}\Sigma[k,E_0(k)],
\label{eq:onshell}
\end{equation}
which does not involve solving the Dyson integral equation. This would
define an on-shell graphene velocity given by $E(k) = v_2^*(k) k$. 
Note that the imaginary part of on-shell self energy near Fermi
surface behaves as Im[$\Sigma[k,E_0(k)] \propto [E_0(k) - \mu]^2 \ln
  [E_0(k) - \mu]$, where $\mu = E_0(k_F)$. \cite{dassarma2007} 
Thus the on-shell imaginary part of the 
self-energy certainly vanishes as $k \rightarrow k_F$.
In fact, the inelastic scattering rate, an extensively used physical
quantity of much experimental interest, is precisely the on-shell
approximation to the Im[$\Sigma(k,E)]$. 
We will compare the on-shell and the off-shell graphene self-energy
solution within our dynamical RPA scheme because the issue of which of
these two is a better approximation to the Coulomb self-energy problem
has been much discussed in the literature over the last 50 years
\cite{dubois1959,rice1965,ting1975,vinter1975,zhang2005}. 
These past discussions of course focused only on regular nonchiral
parabolic electron systems with only intraband contributions and our
current work is on chiral, linearly dispersing, gapless graphene carriers
where both intra- and inter-band contributions to the self-energy must
be included in the theory (as we do in our current work).
We note that there is no deep principle that could make the "correct"
theoretical choice between the on-shell and the off-shell
approximation within the $G_0W$ self-energy scheme since both schemes
obey the Ward identity and the Dyson equation up to the order in the
interaction the theory is meant to be valid.  
(i.e., leading order in the dynamically screened interaction $W$).
The only effective
choice between these approximations is an operational one based on the
detailed quantitative comparison with the experimental data as has
already been discussed in the past
\cite{rice1965,ting1975,vinter1975,zhang2005}.

It may be worthwhile for us to make some remarks on the on-shell
versus off-shell many-body self-energy approximation in the context of
graphene physics, and why this may be a question of substantial
importance. This issue arises because the dynamical Coulomb
self-energy can only be calculated approximately in the leading-order
GW-RPA approximation. If an exact self-energy function $\Sigma(k,E)$
is available, then obviously the full Dyson equation solution would be
necessary to obtain the quasiparticle energy dispersion $E(k)$, making
the off-shell velocity the only meaningful quantity. It has, however, been
pointed out \cite{dubois1959,rice1965,ting1975,vinter1975,zhang2005}
that the leading-order 
screened interaction expansion 
(i.e., the ring diagram approximation) for the calculated RPA
self-energy necessitates using the leading-order iterative solution of
$\Sigma(k,E)$, i.e., using the on-shell approximation
$\Sigma(k,E_0(k))$, instead of the full solution of the integral Dyson
equation so as not to mix various orders of approximation in the
theory. In particular, it has been argued\cite{rice1965} that the on-shell
approximation of the self-energy is, in fact, quantitatively better
than the off-shell approximation within
the GW-RPA scheme of the Coulomb self-energy evaluation for metallic
electrons. The subject has been controversial
\cite{ting1975,zhang2005,vinter1975} with no definite 
consensus in the literature in spite of the almost 50-year history of
the topic. Our careful quantitative comparison of on-shell and
off-shell approximations, when compared with the available
experimental data in graphene should lead to a resolution of this old
and important theoretical controversy.
Our results tend to indicate that the off-shell
approximation works better for graphene as described below.

We emphasize, so that there is no misunderstanding of this point, that
if the self-energy is evaluated exactly then one must always solve the
Dyson equation, and as such the off-shell theory defined by
Eq.~(\ref{eq:dyson}) is the only option for the quasiparticle dispersion
in any exact theory. The question arises, however, whether solving the
Dyson equation is always preferable for obtaining the quasiparticle
energy dispersion $E(k)$ even when the self-energy has been calculated
in an approximate theory. In particular, if the self-energy is
calculated in the leading-order expansion in the dynamically screened
Coulomb interaction, as it is in the current RPA-ring diagram $GW$
theory and in all the existing theories in the literature, then it is
unclear which approximation, on-shell or off-shell, is a better
quantitative approximation.

Iterating Eq.~(\ref{eq:dyson}) formally order by order, i.e., solving
the implicit integral equation for the off-shell self-energy defined
by Eq.~(\ref{eq:dyson}) through the standard iterative technique, we
get
\begin{eqnarray*}
E_1(k) & = & E_0(k) + \Sigma(k,E_0(k)) \\
E_2(k) & = & E_0(k) + \Sigma(k,E_0(k)+\Sigma(k,E_0(k))) \\
E_3(k) & = & E_0(k) + \Sigma(k,E_0(k)+\Sigma(k,E_0(k)+\Sigma(k,E_0(k)))) \\
E_4(k) & = & ...
\end{eqnarray*}
We note that the leading-order iterative solution above is the same as
the on-shell approximation defined in Eq.~(\ref{eq:onshell}). We now
note that the $GW$ approximation used in the current (and many other)
work involves (see Fig.~\ref{fig:fig1}) calculating the self-energy
formally as a leading-order expression in the screened Coulomb
interaction $\Sigma \sim W$, and as such all iterations of the
self-energy beyond the on-shell approximation must necessarily involve
terms which are formally higher order in $W$, which would be
inconsistent with the leading order $GW$ calculation. This is in fact
the argument behind the claim that the on-shell approximation may be a
better approximation than the off-shell one within the leading-order
$GW$ theory. Within this leading-order theory there is no obvious way to
claim that the full solution of the Dyson equation (i.e. the off-shell
approximation) is necessary better than the leading-order solution of
the Dyson equation (i.e. the on-shell approximation) since the former
necessarily includes only an uncontrolled subset of higher-order terms
in the screened interaction whereas the latter is manifestly
leading-order in the screened interaction everywhere. We therefore
provide results for both on-shell and off-shell approximations for
future comparison with experiments.

\section{Renormalized Fermi velocity}

For a given self-energy we can calculate the Fermi velocity by
differentiating the self-energy with respect to the wave vector at $k=k_F$, that
is, $v^*(k) = dE(k)/dk|_{k=k_F}$.
The renormalized Fermi velocities $v_F^*$ corresponding to the
appropriate self-energies of Eqs.~(2) and (3), respectively, are given by
\begin{equation}
\frac{v_1^*(k)}{v_0} = \frac{1 + \frac{1}{v_0}\frac{\partial}{\partial
    k} {\rm Re}\Sigma(k,\omega)|_{\omega=E(k)}}
     {1-\frac{\partial}{\partial \omega} {\rm
         Re}\Sigma(k,\omega)|_{\omega=E(k)}},
\label{eq:voff}
\end{equation}
\begin{equation}
\frac{v_2^*(k)}{v_0} = 1+\frac{1}{v_0}\frac{d}{dk}{\rm Re}
\Sigma[k,E_0(k)],
\label{eq:von}
\end{equation}
and $v_{F_1}^*=v_1^*(k=k_F)$, $v_{F_2}^* = v_2^*(k=k_F)$. 
In our calculation of the on-shell velocity at the Fermi surface (
$k=k_F$ and $E_0(k) = E_F$), we take the limit of both $k$ and $\omega =
E_0(k)$ going to the Fermi surface simultaneously. 
Along this path the imaginary part of the self energy
becomes identically zero at the Fermi surface. However, taking $k$ and
$E_0(k)$ to the Fermi surface independently gives rise to the
different results and would be incorrect. We note
that, while $v_F^*$ (both for on-shell and off-shell approximations)
depends on the coupling constant (i.e., the background dielectric
constant $\kappa$) and the carrier density ($n$), the general
renormalizied graphene velocity $v^*(k)$ depends also on the momentum
$k$, implying that the interacting quasiparticle graphene energy
dispersion $E(k)$ is no longer simply linear in $k$ (as the
noninteracting energy $E_0=v_0k$ is) even for a given substrate and
fixed density. Thus, we have $v^*(k) \equiv v^*(k;n,\kappa)$;
$v_F^*\equiv v_F^*(n,\kappa)$, and depending on the approximation
scheme (off-shell or on-shell) we have two distinct renormalized
velocities $v_{1,2}^*$. A main goal of our work is to obtain the
($k,n;\kappa$) dependence of the graphene quasiparticle velocity.

\begin{figure}[t]
	\centering
	\includegraphics[width=1.\columnwidth]{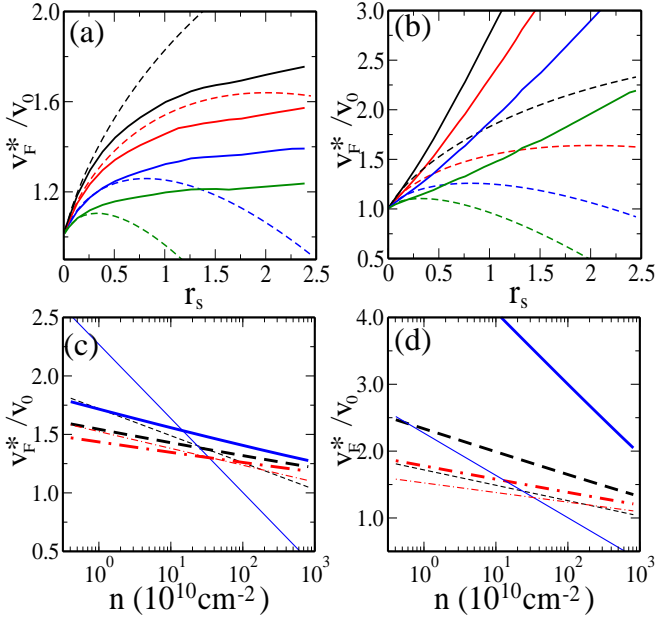}
	\caption{ 
Calculated renormalized Fermi velocity (a) within the off-shell
approximation, $v^*_{F_1}/v_0$, and (b) within the on-shell approximation,
$v^*_{F_2}/v_0$, as a function
          of $r_s$ for several different densities, $n=10^{10}$,
          $10^{11}$, $10^{12}$, and $10^{13}$ cm$^{-2}$ (from top to
          bottom). The dashed lines 
          represent the asymptotic results for small $r_s$ given by
          Eq.~(\ref{eq:fermi}). 
In (c) and (d) the renormalized Fermi velocityies (thick lines) within the off-shell
approximation and within the on-shell approximation are shown, respectively, as a function
          of carrier density for several different $r_s$, $r_s = 0.5$
          (dod-dashed lines), $r_s = 0.8$ (dashed lines), and $r_s =
          2.2$ (solid lines).
          The thin lines 
          represent the asymptotic results for small $r_s$ given by
          Eq.~(\ref{eq:fermi}). 
}
\label{fig:fig2}
\end{figure}

\begin{figure}[t]
	\centering
	\includegraphics[width=1.\columnwidth]{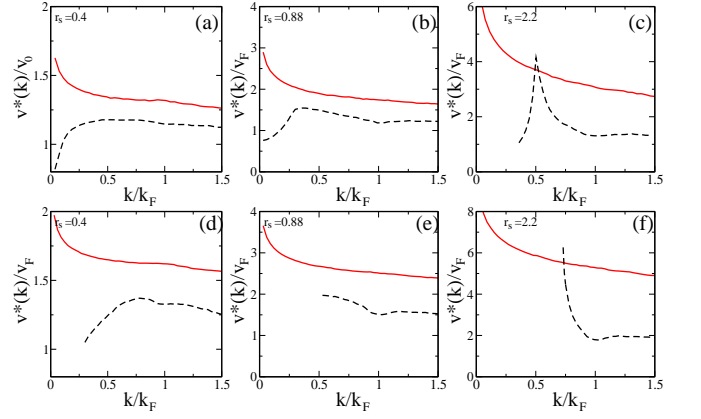}
	\caption{ 
The calculated $v^*(k)$ as a function of $k/k_F$ are shown for (a)
$r_s = 0.4$, (b) $r_s = 0.88$, and $r_s = 2.2$
and a fixed carrier density $n=10^{12}$ cm$^{-2}$. The black dashed
(red solid) line indicates the renormalized Fermi velocity within off-shell
(on-shell) approximation.  In (c) the velocity within off-shell is not
defined for $k/k_F \alt 0.3$ (i.e., there is no solution in Dyson's
equation correpsonding to the quasiparticle.)
In (d), (e), and (f) we show $v^*(k)$ for carrier density $n=10^{10}$
cm$^{-2}$ and $r_s=0.4$, $r_s=0.88$, and $r_s = 2.2$, respectively. Since
there are no solutions in Dyson's equation correponding
to the quasiparticle the renormalized off-shell velocity at small $k$ is not shown
in (d), (e), and (f).
}
\label{fig:fig3}
\end{figure}

Before presenting our full numerical results (Figs.~\ref{fig:fig2} and
\ref{fig:fig3}) which is one of  
our two main results in this work, we give the leading-order analytic
result for the Fermi velocity $v_F^* \equiv v^*(k=k_F)$ for gated
graphene
\begin{equation}
\frac{v_F^*}{v_0} = 1-\frac{r_s}{\pi} \left [ \frac{5}{3} + \ln r_s
  \right ] + \frac{r_s}{4} \ln \frac{k_c}{k_F},
\label{eq:fermi}
\end{equation}
where $r_s \equiv e^2/(v_0 \kappa)$ is the so-called (background
dielectric constant dependent) graphene fine-structure constant which
defines the dimensionless electron-electron interaction strength or
the Coulomb coupling constant for the problem and $k_c \sim 1/a$
(where $a$ is the graphene lattice size) is the
ultraviolet cut-off for the problem. Throughout this paper $k_c =
1/(2.46$\AA) is used. Here $\kappa = (\kappa_s+1)/2$
where $\kappa_s$ is the lattice dielectric constant of the substrate
material. The analytic formula given in Eq.~(\ref{eq:fermi}) has
earlier been derived in the literature \cite{dassarma2007}, and we
provide it here for the sake of completeness and comparison.

It is
important to emphasize that Eq.~(\ref{eq:fermi}) applies
to both on-shell and off-shell approximations as they have exactly the
same leading-order expansions in $r_s$, and Eq.~(\ref{eq:fermi})
provides an exact expression in the $r_s \rightarrow 0$ limit since
the ring diagrams are the most divergent diagrams in this
leading-order (in $r_s$) limit. A curious feature of
Eq.~(\ref{eq:fermi}), which is completely unique to graphene with its
linear massless and gapless energy dispersion, is that the velocity
renormalization depends on two distinct dimensionless coupling
parameters, $r_s$ [$\propto (v_0\kappa)^{-1}$] and $k_c/k_F$ [$\propto
  (\sqrt{n}a)^{-1}$] arising respectively from the Coulomb interaction
strength ($r_s$) and the ultraviolet lattice cut-off ($k_c/k_F$). In
ordinary parabolic electron systems, the Coulomb coupling strength
$r_s$ itself depends on the carrier density (e.g., in 2D parabolic
electron systems with a mass $m$, $r_s \sim m/\sqrt{n}$), and no
ultraviolet cut-off is necessary. Thus, graphene is a very special and
interesting Coulomb system with the interaction strength $r_s$ itself
being density independent, but renormalization effects being
logarithmically density dependent through the ultraviolet
divergence. 
Such logarithmic dependence on the ultraviolet cut-off, while being
common in relativistic field theories, is rare in solid state physics
where the infrared Coulomb divergence is typically the main problem.
We also mention that the second term (the square bracket)
in Eq.~(\ref{eq:fermi}) and the last term [containing $\ln (k_c/k_F)$]
arise respectively from intraband and interband transitions.
Depending on the carrier density and the background dielectric
constant, one or the other contribution could dominate, but in general
both must be kept in the theory for extrinsic graphene.
We note that the ultraviolet logarithmically divergent term in
Eq.~\ref{eq:fermi} goes as $\ln(n/n_c)$ where $n_c$ is the cut-off
density defining $k_c$, and as such the ultraviolet renormalization of
the graphene coupling constant can be directly studied by tuning the
carrier density $n$ with its effect being larger for larger $r_s$.

In Fig.~\ref{fig:fig2} we show our full numerical calculations for the renormalized
graphene velocity in the off-shell [Fig.~\ref{fig:fig2}(a)] and on-shell
[Fig.~\ref{fig:fig2}(b)] approximations as a function of $r_s$ for different
carrier densities ($n$) comparing the results with the analytic
formula given by Eq.~(\ref{eq:fermi}). In Figs.~\ref{fig:fig2}((c) and
(d) the renormalized velocities as a function of density $n$ for
different $r_s$ values are shown.
Several features stand out in
Fig.~\ref{fig:fig2}: ($i$) the velocity renormalization is a strong
function of both $r_s$ and $n$ in both approximation schemes; ($ii$)
the renormalization is substantially stronger in the on-shell than in
the off-shell approximation; ($iii$) the leading-order analytic formula
[Eq.~(\ref{eq:fermi})] applies only for $r_s \alt 0.4-0.8$ depending
on the density and the approximation scheme.

In Fig.~\ref{fig:fig3} we show our
numerically calculated $v^*(k)$ for three different $r_s$ values
as a function of momentum for the fixed carrier density
$n=10^{12}$ cm$^{-2}$ for both approximation schemes. There are
several interesting features of Fig.~\ref{fig:fig3}: ($i$) both
approximations imply weak 
nonlinearity in the renormalized graphene quasiparticle dispersion
except for $k < 0.4 k_F$; ($ii$) the many-body renormalization is
substantially stronger for the on-shell approximation; ($iii$) at
low momentum, $k \alt 0.5k_F$, the off-shell (on-shell)
renormalization indicates nonlinear negative (positive) velocity
corrections in graphene; ($iv$) as $r_s$ increases or at low densities
the renormalized off-shell velocity is strongly nonlinear;
($v$) for large $r_s$ values or at low densities the
renormalized off-shell velocity is not defined at small $k$, that is,
there is no solution in Dyson's equation [see Eq.~(\ref{eq:dyson})]
corresponding to the quasiparticle energy. Thus the quasiparticle
feature at low momentum disappears at high $r_s$ values or at low
densities. We emphasize that some aspects of this physics were already
apparent in Ref.~[\onlinecite{arpes}], but the effect should be much more
pronounced at higher (lower) $r_s$ ($n$) values.

One feature of Figs.~\ref{fig:fig2} and \ref{fig:fig3} requires
particular emphasis since it is quite 
non-obvious: On-shell and off-shell approximations start deviating
from each other substantially already in the rather weak-coupling
situation of $r_s \agt 0.5$. Thus, it is clearly important to know by
comparing with experiments which one is the better approximation.

\begin{figure}[t]
	\centering
	\includegraphics[width=1.\columnwidth]{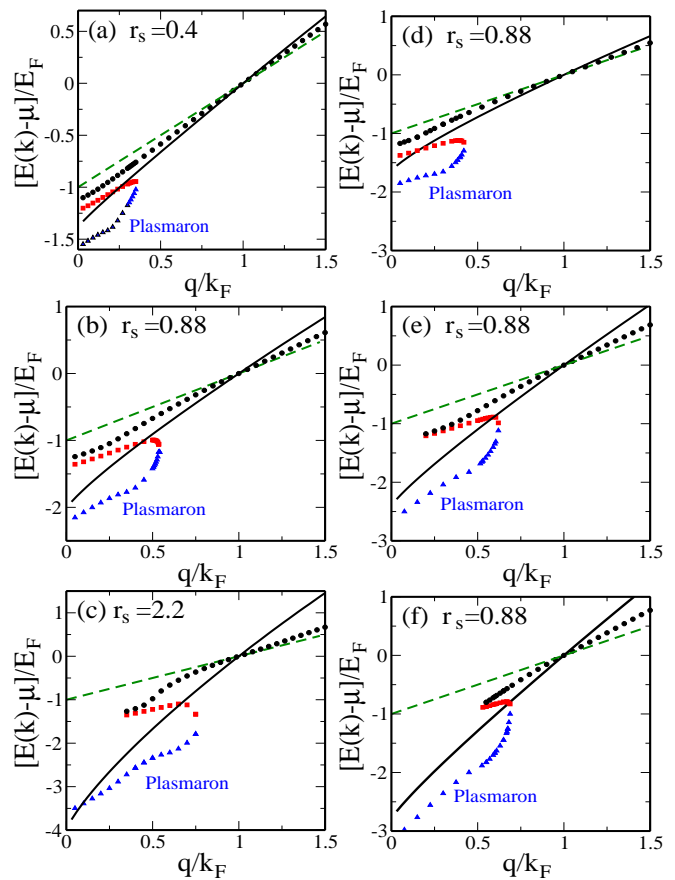}
	\caption{ 
The renormalized quasiparticle energy dispersions as a function of wave vector 
for the electron density $n=10^{12}$ cm$^{-2}$ and 
for various $r_s$, (a) $r_s=0.4$, (b)
$r_s=0.88$, and (c) $r_s=2.2$. In (d)(e)(f)
the renormalized quasiparticle energy dispersions are shown as a function of wave vector 
for a fixed $r_s=0.88$ and for different electron densities (d) $n=10^{13}$ cm$^{-2}$,  
(e) $n=10^{11}$ cm$^{-2}$, and (f) $n=10^{10}$ cm$^{-2}$.  
Green dashed (black solid) lines represent the non-interacting 
(on-shell) energy dispersion, and symbols are the solutions of the
Dyson's integral equation. The dots (triangles) represent the
dispersion of off-shell 
quasiparticle mode (plasmaron modes). The squares indicate the
dispersion of the overdamped mode, which has insignificant spectral weight.
}
\label{fig:fig4}
\end{figure}

\section{quasiparticle dispersion and spectral function}

Although Figs.~\ref{fig:fig2} and \ref{fig:fig3} present one of our
main quantitative theoretical results 
showing the density, coupling constant, and momentum dependence of the
renormalized graphene quasiparticle velocity, we now provide some
calculated results for the many-body quasiparticle dispersion and
spectral function since the quasiparticle velocity is derived from the
quasiparticle energy dispersion [c.f. $E(k) = v^*(k)k$] which is
contained in the interacting spectral function of the system. 
Even though we have used the well known many body diagrammatic
perturbation theory ($G_0W$ approximation) we find very interesting
and experimentally relevant
results in the graphene energy dispersion which have not been discussed in
the existing literature.
These results are quite unanticipated.

\begin{figure}[t]
	\centering
	\includegraphics[width=1.\columnwidth]{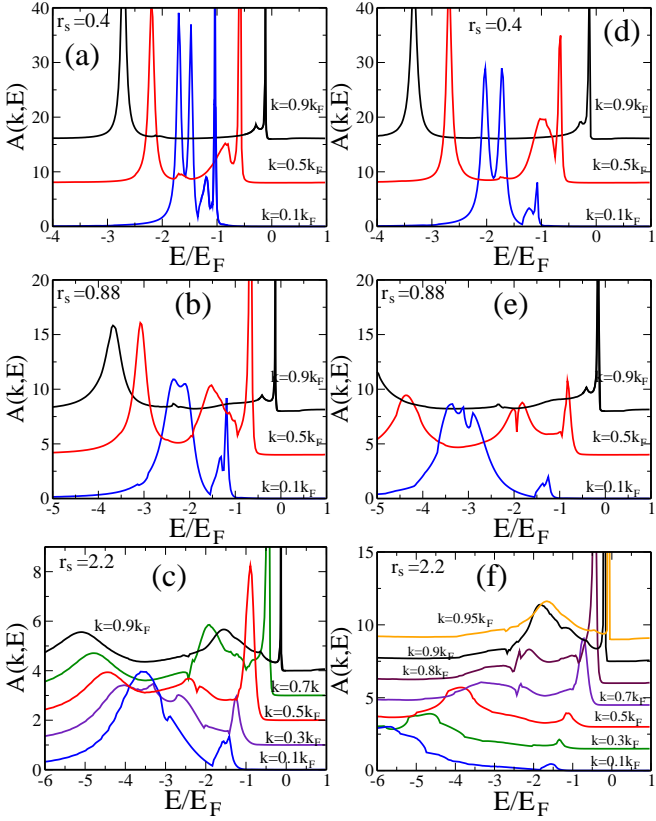}
	\caption{ 
Calculated quasiparticle spectral function $A(k,E)$ as a function of energy
for different wave 
vectors and three different $r_s$ values. The lines for different wave vectors
are shifted upward for clarity.
In (a), (b), and (c) the density $n=10^{12}cm^{-2}$ is used,
and the density $10^{10}cm^{-2}$ is used in
(d), (e), and (f).
}
\label{fig:fig5}
\end{figure}

In Fig.~\ref{fig:fig4}, we show our calculated quasiparticle energy
dispersion for different densities and $r_s$ values. In (a), (b), and
(c) results for a fixed density 
$n=10^{12}$ cm$^{-2}$ and for different $r_s$ values ($r_s=0.4$, 0.88,
and 2.2 respectively) are shown, and in (d), (e), and (f) results for a fixed
$r_s=0.88$ and for different densities ($n=10^{13}$, $n=10^{11}$
and $n=10^{10}$ cm$^{-2}$, respectively) are shown in 
both on-shell and off-shell approximations comparing them with the
noninteracting energy dispersion $E_0(k) = v_0 k$. The Fermi velocity
in each case is given by the $k=k_F$ point. 
We note that $r_s=0.4$, 0.88, 2.2 correspond to graphene on SiC (or
BN), SiO$_2$, vacuum, respectively.
The important thing to
note is that the on-shell approximation indicates very strong
quasiparticle dispersion renormalization, particularly for suspended
graphene ($r_s=2.2$), which may be unphysical, thus perhaps ruling out
the validity of the on-shell approximation. Consistent with the
results presented in Fig.~\ref{fig:fig2}, the off-shell
approximation predicts rather small many-body dispersion (and
velocity) renormalization, but it does make a very dramatic prediction
about the complete failure of the quasiparticle picture for small
momentum in suspended graphene ($r_s=2.2$). In particular, the
off-shell quasiparticle mode for $r_s = 2.2$ (and $n=10^{12}$
cm$^{-2}$) completely disappears for $k<0.4k_F$ (we have checked the
same to be true for other carrier densities also), leaving only a weak
plasmaron mode which shows up for all $r_s$ values at low wave vectors.
Thus a
clear prediction of our theory is that the quasiparticle dispersion in
suspended graphene ($r_s=2.2$) will look dramatically different for
$k\alt 0.5 k_F$ with the complete disappearance of any particle-like
feature in the graphene dispersion. 
As shown in Fig.~\ref{fig:fig4}(f) we find that the similar feature
(i.e. the complete disappearance of a particle-like feature)
develops as the density decreases. For $r_s = 0.88$ and high enough
densities [$n > 10^{11} cm^{-2}$, see Fig.~\ref{fig:fig4}(b) and (d)]
the quasiparticle-like dispersion is well defined even at $q=0$, but
for $n \alt 10^{11} cm^{-2}$ the quasiparticle feature starts to
disappear near $q=0$. Thus, there are two distinct ways of approaching
'strong-coupling' in graphene either by going to smaller $k_F$
(i.e. lower density) keeping $r_s$ fixed or by keeping $k_F$ (or
density) fixed and increasing $r_s$. However, these two
strong-coupling approaches give 
non-equivalent results as shown in Fig.~\ref{fig:fig4}. 
Note that in ordinary electron 2D gas only changing $r_s$ is allowed
to approach the strong coupling limit
whereas in graphene one has one more adjustable parameter to approach
the strong coupling limit, i.e., by changing $r_s$ through background
dielectric constant of substrate and by changing $k_F$ through the
back gate voltage.
Since some aspects of this interesting physics were already observed
in Ref.~[\onlinecite{arpes}] we believe that experiments at higher
(lower) $r_s$ ($n$) values should lead to a striking confirmation of
our theoretical predictions.

This prediction can be directly
tested in ARPES \cite{arpes,arpes1,arpes2} and STS \cite{luican2011,sts,sts2}
measurements. 
Earlier theoretical work\cite{polini2008,hwang2008} on the graphene
spectral function completely missed this bizarre collapse of the
quasiparticle picture for suspended graphene since they
\cite{polini2008,hwang2008} focused on graphene on substrates. We note
that $r_s=2.2$ is a relatively weak-coupling regime compared with 3D
metals ($r_s \sim 3-6$) and 2D semiconductors ($r_s \sim 5-30$) where
no such dispersion instability in the interacting spectral function
has ever been predicted.

\begin{figure}[t]
	\centering
	\includegraphics[width=1.0\columnwidth]{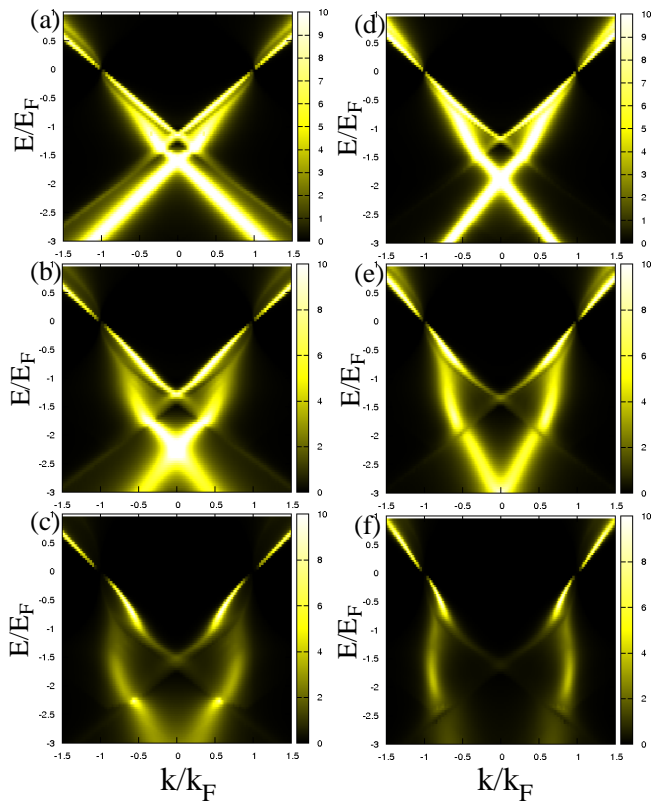}
	\caption{ 
Density plot of the quasiparticle spectral function $A(k,E)$
as a function of energy $E/E_F$ and wave 
vectors $k/k_F$ for three different $r_s$ values, (a) and (d)
$r_s=0.4$, (b) and (e) $r_s=0.88$, (c) and (f) $r_s=2.2$. 
In (a), (b), and (c) the density $n=10^{12}cm^{-2}$ is used,
and the density $10^{10}cm^{-2}$ is used in
(d), (e), and (f).
}
\label{fig:fig6}
\end{figure}

To further emphasize the low wave vector interaction-driven instability
in the graphene quasiparticle spectrum discussed above, we show in
Fig.~\ref{fig:fig5} our calculated interacting graphene spectral function given by
$A(k,E) \equiv - 2{\rm Im} G(k,E)$, which directly provides the
spectral strength of $E(k)$ features in the graphene spectrum with the
noninteracting result being precisely $A_0(k,E) = 2\pi
\delta[E-E_0(k)]$. 
The calculated spectral function in Fig.~\ref{fig:fig5} clearly 
shows that for $r_s=2.2$, the single particle-like feature disappears
completely from the graphene spectral function for $k < 0.5 k_F$ ($0.7
k_F$) for $n=10^{12}$ ($10^{10}$) cm$^{-2}$, and the spectral weight
basically gets distributed broadly over the incoherent background. For
$r_s=0.4$ (graphene on SiC or BN), however, some sharp spectral
features persist down to low wave vectors, whereas graphene on SiO$_2$
($r_s = 0.88$) is intermediate in nature with sharp particle-like
spectral features being present in the spectrum at low wave vectors
and at high density, but disappearing at low density. 
Our theory, which has
direct observable consequences for graphene experiments, thus predicts
that the whole linear dispersing noninteracting energy band picture in
graphene should completely collapse in suspended graphene for $k \alt
0.5 k_F$ in the $10^{10} - 10^{12}$ cm$^{-2}$ density range where all
the spectral weight should disappear into a broad incoherent
background due to electron-electron interaction.
We emphasize that earlier graphene spectral function calculations
\cite{polini2008,hwang2008} completely missed this dramatic behavior
arising from the linear chiral nature of graphene.

To clearly see the disappearance (instability) of the graphene energy
dispersion at 
small wave vectors (near Dirac point),
in Fig.~\ref{fig:fig6} we show the calculated spectral function, $A(k,E)$, as a
color plot in $k/k_F$ and $E/E_F$ space
for two different densities (a),(b),(c) $n = 10^{12} cm^{−2}$ and
(d),(e),(f) $n= 10^{10} cm^{−2}$, respectively. 
In this figure the narrow bright area indicates the highest
spectral weight.
For $r_s=0.4$ [(a) and (d)] the single particle-like linear dispersion
has the most spectral weight and well defined plasmaron features
appear below the single particle-like peak.
However for $r_s=2.2$ [(c) and (f)] there are no well defined peaks
at small wave vectors and the spectral weight
is distributed broadly over the incoherent background. The broadening
of spectral weight at small wave vectors is more pronounced at low
density (f). The disappearance (instability) develops gradually as the
wave vector decreases, and 
the anomalous dispersion is stronger for lower density and higher
$r_s$. 
We suggest that spectral function measurements via ARPES or STS be
carried out on suspended graphene to directly observe our predicted
instability.

\section{conclusion}

To conclude, we have provided the detailed quantitative results for
nonlinear interacting
graphene dispersion as a function of density, coupling constant, and
momentum, predicting in the process a spectacular collapse of the
quasiparticle picture in suspended graphene for $k \alt
0.5k_F$, which can be directly tested in experiments.
Our results indicate that many-body
renormalization of graphene velocity is possible up to a factor of two
as a function of
either carrier density or wave vector if a large range of density and
wave vector can be explored in future experiments. 
Our work is a generalization of earlier works
\cite{polini2008,hwang2008,arpes} in the literature to large $r_s$
values and lower carrier densities where the anomalous dispersion
properties of graphene should lead to a striking collapse of the
quasiparticle picture as shown in our results.

Even though we have used the well known many body diagrammatic
perturbation theory ($G_0W$ approximation) our main findings (i.e., Fermi
velocity renormalization in both on-shell and off-shell approximations
and an instability in the graphene energy dispersion) are totally
unexpected. Especially, the disappearance of the quasiparticle
features at small wave vectors as
the interaction strength increases is totally unanticipated and
has not earlier been predicted or discussed anywhere in the existing
literature to the best of our knowledge. Since the current available experiments
\cite{arpes,arpes1,arpes2,sts,sts2,luican2011} do
not indicate any signature of this instability 
we believe that this work is important and of interest because the
results presented in this paper are relevant for future experiments.  
The current available experiments have mostly been performed at small $r_s$
values for graphene on substrates. So, to see our prediction of the
instability it is required to do 
experiments with samples having small electron density and high $r_s$
value such as suspended graphene.   
Since there are two distinct ways of approaching
'strong-coupling' in graphene either 
by keeping density $n$ fixed and increasing $r_s$ or
by going to smaller $n$ keeping $r_s$ fixed,
the similar many body featuren may appear as the density decreases for
a fixed $r_s$ value. Thus, in 
stead of increasing the 
$r_s$ value we can measure these many body effects by reducing density in
graphene on SiO$_2$ or SiC substates.

\section{acknowledgments}
This work is supported by US-ONR.

\end{document}